\pdfoutput=1
\documentclass[twocolumn,trackchanges]{aastex701}
\usepackage{amsmath}
\usepackage{breqn}

\newcommand{\Cloudy}{\textsc{Cloudy}}

\font\manual=manfnt at 7pt \def\dbend{\hbox{\raise0.9ex\hbox{\manual\char127\hspace{0.6em}}}}

\providecommand{\e}[1]{\ensuremath{\times 10^{#1}}}

\newcounter{INTERNALionstage}

\def\gtsim{\mathrel{\hbox{\rlap{\hbox{\lower4pt\hbox{$\sim$}}}\hbox{$>$}}}}
\def\lesssim{\mathrel{\hbox{\rlap{\hbox{\lower4pt\hbox{$\sim$}}}\hbox{$<$}}}}

\def\A{{\rm\thinspace \AA}}

\def\g{{\rm\thinspace g}}

\def\K{{\rm\thinspace K}}

\def\m{{\rm\thinspace m}}

\def\s{{\rm\thinspace s}}

\def\W{{\rm\thinspace W}}

\def\fexxv{\mbox{{\rm Fe~{\sc xxv}}}}

\def\h0{\mbox{{\rm H}$^0$}}
\DeclareMathAlphabet{\vib}{OML}{cmm}{m}{it}

\begin{document}

\title{Complex Ionization and Velocity Structures in GX~340+0 X-ray Binary Revealed by XRISM}

\author[0000-0002-4469-2518]{Priyanka Chakraborty}
\affiliation{Center for Astrophysics $\vert$ Harvard \& Smithsonian, 60 Garden Street, Cambridge, MA, 02138}
\email{priyanka.chakraborty@cfa.harvard.edu}

\author[0000-0003-4284-4167]{Randall 
Smith}
\affiliation{Center for Astrophysics $\vert$ Harvard \& Smithsonian, 60 Garden Street, Cambridge, MA, 02138}
\email{}

\author[0000-0002-5466-3817]{Lia Corrales}
\affiliation{Astronomy Department, University of Michigan, Ann Arbor, MI 48109, USA}
\email{}
 
\author[0000-0001-8470-749X]{Elisa Costantini}
\affiliation{SRON Netherlands Institute for Space Research, Leiden, The Netherlands}
\email{}

\author[]{Mar{\'i}a D{\'i}az Trigo}
\affiliation{ESO, Karl-Schwarzschild-Strasse 2, 85748, Garching bei M{\''u}nchen, Germany}
\email{}

\author[0000-0003-3462-8886]{Adam Foster}
\affiliation{Center for Astrophysics $\vert$ Harvard \& Smithsonian, 60 Garden Street, Cambridge, MA, 02138}
\email{}

\author[0000-0001-9464-4103]{Caroline Kilbourne}
\affiliation{NASA / Goddard Space Flight Center, Greenbelt, MD 20771, USA}
\email{}

\author[0000-0002-8961-939X]{Renee Ludlam}
\affiliation{Department of Physics \& Astronomy, Wayne State University, 666 West Hancock Street, Detroit, MI 48201, USA}
\email{}

\author[0000-0002-6660-9375]{Takao
Nakagawa}
\affiliation{Institute of Space and Astronautical Science (ISAS),
Japan Aerospace Exploration Agency (JAXA),
Sagamihara, 252-5210 Kanagawa, Japan}
\affiliation{Advanced Research Laboratories, Tokyo City University, 1-28-1
Tamazutsumi, Setagaya, 158-8557 Tokyo, Japan}
\email{}

\author[0000-0002-6374-1119]{Frederick S. Porter}
\affiliation{NASA / Goddard Space Flight Center, Greenbelt, MD 20771, USA}
\email{}

\author[0000-0002-1049-3182]{Ioanna
Psaradaki}
\affiliation{MIT Kavli Institute for Astrophysics and Space Research, Massachusetts Institute of Technology, Cambridge, MA 02139, USA}
\email{}

\author[0000-0001-6314-5897]{Hiromitsu Takahashi}
\affiliation{Department of Physics, Hiroshima University, Hiroshima 739-8526, Japan}
\email{}

\author[0000-0001-6314-5897]{Tahir Yaqoob}
\affiliation{NASA / Goddard Space Flight Center, Greenbelt, MD 20771, USA}
\affiliation{Center for Research and Exploration in Space Science and Technology, NASA/GSFC (CRESST II), Greenbelt, MD 20771, USA}
\affiliation{Center for Space Science and Technology, University of Maryland, Baltimore County (UMBC), 1000 Hilltop Circle, Baltimore, MD 21250, USA}
\email{}

\author[0000-0002-8163-8852]{Sascha Zeegers}
\affiliation{European Space Agency (ESA), European Space Research and Technology Centre (ESTEC), Keplerlaan 1, 2201 AZ Noordwijk, The
Netherlands}
\email{}




\begin{abstract}
{We present the first high-resolution XRISM spectrum of the neutron star low-mass X-ray binary GX~340+0, revealing unprecedented detail in its emission and absorption features.  The spectrum reveals a rich and complex Fe~\textsc{xxv} He$\alpha$ line profile and a P-Cygni profile from Ca~\textsc{xx}. 
We use the state-of-the-art spectral synthesis code $\Cloudy$ to model the emission and absorption features in detail.} Our analysis reveals multi-ionization and multi-velocity structures, where the combination of broad (\(\sim 800 \, \text{km~s$^{-1}$}\)) and narrow (\(\sim 360 \, \text{km~s$^{-1}$}\)) line components, along with rest-frame and blueshifted emission and absorption lines, accounts for the observed line profile complexity. 
We identify a modest $\sim$ 2735 km~s$^{-1}$ accretion disk wind exhibiting both absorption and emission features.
We also detect a relativistic reflection feature in the spectrum, which we model using \texttt{relxillNS}—specifically designed to characterize X-ray reprocessing in accretion disks around neutron stars.
Furthermore, we examine the detailed physics of the Fe~\textsc{xxv} He\(\alpha\) complex, focusing on the forbidden-to-resonance line ratio under the influence of continuum pumping
and optical depth effects.
\end{abstract}
\keywords{X-rays---binaries, techniques---spectroscopic, stars---neutron}


\section{Introduction} 

GX 340+0 has long been recognized as a persistently bright low-mass X-ray binary, with a neutron star accretor, as identified from Aerobee rocket data \citep{1971ApJ...169L..45M}. It was classified as a “Z” source 
based on its X-ray color-color diagram \citep{1989A&A...225...79H}. 
Spectral analysis of GX 340+0 using EXOSAT observations first revealed the complex nature of its spectrum \citep{1993A&A...273..123S}. 
Although the radio and infrared counterparts of the target were soon identified \citep{1993A&A...267...92P, 1993AJ....106...28M}, its optical counterpart is still unknown. 

\vspace{0.02in}
\noindent GX 340+0 has been extensively studied with various X-ray observatories, leading to a range of interpretations, including some conflicting views, regarding its emission and absorption features, particularly the Fe K$\alpha$ lines. The first broadband spectrum of GX 340+0 in the range 0.1 to 200 keV was obtained by BeppoSAX \citep{2004NuPhS.132..616L}. 
They detected a broad Gaussian Fe emission line  at 6.75 keV for the first time, which was confirmed in \citet{2005ApJ...620..274U}.

\vspace{0.02in}
\noindent Observations with \textit{XMM-Newton} later revealed that the broad Fe K$\alpha$ line exhibits an asymmetric profile \citep{2009ApJ...693L...1D}, consistent with a relativistically smeared emission originating from the reflection of photons off the accretion disk. They inferred that recombination of highly ionized He-like iron produces the emission line, which is then deformed and broadened by Doppler and relativistic effects.  \citet{2010ApJ...720..205C} also reported an asymmetric 
line profile, which could be fit equally well with a relativistic emission line profile with variable Fe abundance, as well as  an Fe~\textsc{xxvi} absorption line superposed on a Fe~\textsc{xxv} emission line.

\vspace{0.02in}
\noindent \citet{2010A&A...522A..96N} analyzed archival \textit{XMM-Newton} observations of 16 neutron star low-mass X-ray binaries (NS LMXBs), including GX 340+0, and reported highly symmetric iron line profiles, concluding that the lines are most likely emitted from the Accretion Disk Corona (ADC).
 In contrast to earlier studies, \citet{2016ApJ...822L..18M} presented evidence of ionized absorption features in the Fe K band. Their analysis of \textit{Chandra}/HETG observational data revealed a prominent absorption feature at 6.9 keV, along with several weaker absorption lines. They interpreted the primary absorption feature as most likely originating from the Fe~\textsc{xxv} He$\alpha$ resonance line (6.7 keV), indicative of a high-speed accretion disk wind. Additionally, they considered the less likely possibility that the feature could correspond to the H-like Fe~\textsc{xxvi} Ly$\alpha$ resonance line (6.97 keV), potentially associated with a modest inflow.

\noindent With high-resolution spectroscopy, we are able to examine the Fe K feature with unprecedented detail.
 \textit{XRISM}/Resolve observations of Cygnus X-3 demonstrated XRISM's transformative capability by resolving the Fe\,\textsc{xxv} He$\alpha$ and Fe\,\textsc{xxvi} Ly$\alpha$ complexes into fine-structure transitions and identifying multiple kinematic and ionization components in both absorption and emission \citep{2024ApJ...977L..34X}. Their superposition produced complex line profiles, including strong P-Cygni features on resonance lines. Self-consistent photoionization modeling enabled the decomposition of these components and the measurement of their individual Doppler velocities.  In the low-mass X-ray binary Circinus X-1, the outflowing photoionized plasma exhibited H-like and He-like lines of S, Ca, Ar, and Fe in both emission and absorption, and was successfully modeled using a self-consistent photoionization framework \citep{2025PASJ..tmp...30T}.

\noindent In this paper, we present a comprehensive spectral analysis of the first high-resolution spectrum of GX 340+0 observed by the \textit{XRISM} X-ray Observatory \citep{2021SPIE11444E..22T}. In particular, this study focuses on the emission and absorption features in the Fe and Ca line complexes. Beyond this, we analyze all identified emission and absorption lines to gain a broader understanding of the physical and dynamical processes influencing GX 340+0. To further interpret these features, we perform detailed photoionization modeling with the state-of-the-art spectral synthesis code $\Cloudy$ \citep{2017RMxAA..53..385F, 2023RMxAA..59..327C}, investigating the mechanisms driving line emission, disk wind properties, and the potential gas dynamics within the system. Furthermore, we identify a relativistic reflection component, which we model using the \texttt{relxillNS} \citep{2022ApJ...926...13G} framework developed for accreting neutron stars, yielding constraints on the ionization state and geometry of the inner accretion disk.

With \textit{XRISM}/Resolve's unprecedented spectral resolution in the Fe K band, we are able to directly observe radiative transfer effects, such as continuum pumping and distinguish between different line formation processes \citep{2021ApJ...912...26C, 2022ApJ...935...70C}, in X-rays.
Line formation processes can be categorized into four distinct cases based on the optical depth and the presence of external radiation- Case A, Case B, Case C, Case D. Case A and Case B occur when no external radiation source is present. In Case A, the Lyman line optical depths are very small, and line photons escape the optically thin cloud without any scattering \citep{1937ApJ....85..330M}. In Case B, the Lyman line optical depths are large enough for photons to undergo multiple scatterings before escaping the cloud \citep{1938ApJ....88...52B}. Case C occurs when an external radiation source illuminates an optically thin cloud \citep{1999PASP..111.1524F}. In this scenario, Lyman lines are enhanced through induced radiative excitation of atoms or ions by continuum photons, a process known as continuum pumping. Case D occurs in the presence of an external radiation source when the cloud is optically thick \citep{2009ApJ...691.1712L, 2016RMxAA..52..419P}. Here, the line intensities are shaped by the competition between continuum pumping and absorption or scattering processes.

Continuum pumping, central to Cases C and D, and typical for binary environments, involves line photons from an external radiation source being absorbed by a plasma, triggering radiative cascades that selectively enhance specific emission lines, such as those within the Fe XXV He$\alpha$ complex, particularly resonance lines with large oscillator strengths \citep{2021ApJ...912...26C}.  Unless the continuum source exhibits strong Lyman absorption lines, continuum pumping will always enhance these resonance emission lines \citep{1999PASP..111.1524F}.
This process minimally affects other lines and serves as a key mechanism differentiating Cases C and D from Cases A and B, which occur in the absence of external radiation.   In this paper, we also discuss the Fe K complex in the framework of Case C and Case D to understand the mechanisms driving its distinctive spectral properties.

This paper is organized as follows: Section 2 details  the observed data reduction and analysis performed for the \textit{XRISM} observations. In Section 3, we detail the photoionization modeling performed using $\Cloudy$, including separate analyses of the continuum, emission, and absorption features alongside reflection modeling with \texttt{relxillNS}. Section 4 studies the detailed physics of the Fe XXV He$\alpha$ complex. Section 5 discusses our results.

\begin{figure}
\centering
\includegraphics[width=0.5\textwidth]{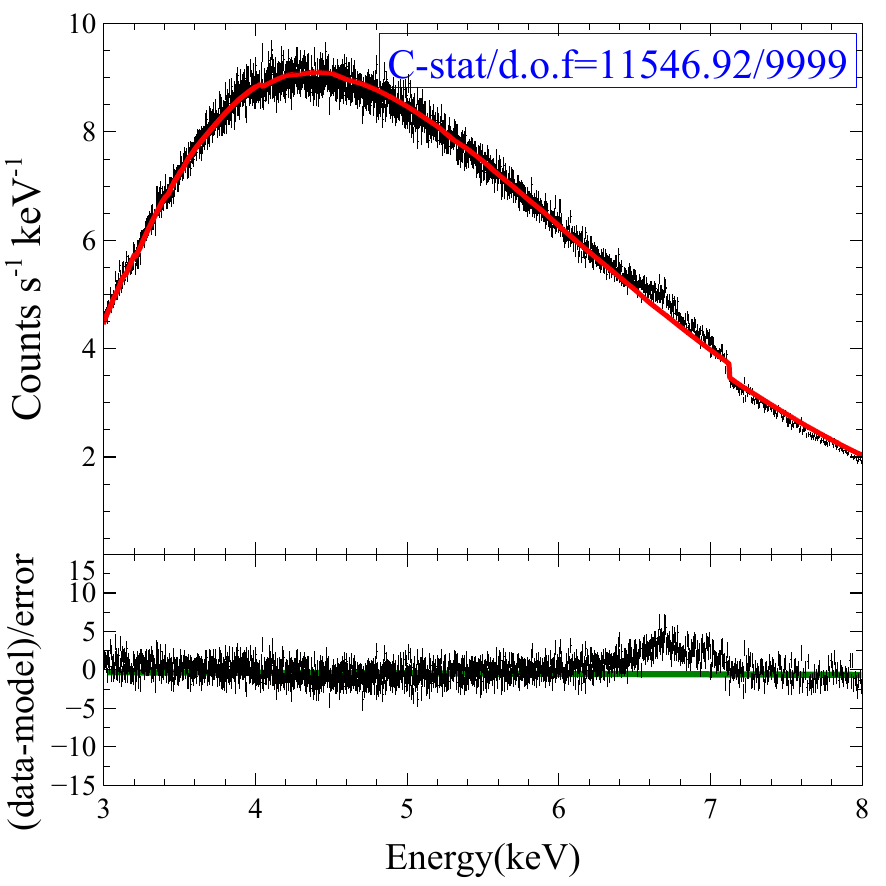}
\caption{The 3--8 keV \textit{XRISM}/Resolve spectrum of GX340+0 overlaid with best-fitting continuum model, significant emission residual is present
between 6--7 keV.
\label{f:fig1}}
\end{figure}

\section{Observations and Data analysis}

 GX 340+0 was observed twice during its performance verification phase.  The first observation, conducted on 2024 Aug 17 with an exposure time of  152.33 ks (Observation ID: 300002010), was followed by a second observation on 2024 Aug 31 with an exposure time of 152.97 ks (Observation ID: 300002020). This study focuses on a time-averaged  analysis of Observation 2\footnote{GX 340+0 is a time-variable source, exhibiting transitions between the normal branch (NB) and flaring branch (FB) during observation 2. However, the signal-to-noise ratio within the individual spectral segments is insufficient to support a detailed, branch-resolved analysis. A time-resolved or Z-track-resolved study remains beyond the scope of this work, but would be a valuable objective for future observations with longer exposure time.}, whereas Observation 1 has been previously analyzed by \citet{2025arXiv250706289L}.

In this analysis, we exclusively utilize observations obtained with the Resolve instrument. The data were reprocessed using Heasoft-6.34, incorporating calibration files from CalDB v8 and adhering to the standard filtering procedures outlined in the XRISM analyses of N132D  and Cyg X-3 \citep{xrismcollaboration2024xrismlightobservationvelocity, xrismcollaboration2024xrismresolveviewfek}. The coordinates of right ascension and declination were constrained to the standard positional coordinates specified for Observation ID  300002020(RA=251.4498, DEC=-45.6110).

 The time dependent energy scale for Resolve was calibrated during the observation using onboard calibration sources that were placed into the field of view (FOV) periodically during earth occultation. The second GX340+0 observation, reported here, contained 11 such fiducial measurements, and the energy scale was corrected using the standard method for Resolve \citep{2016SPIE.9905E..0WP, 2024SPIE13093E..1KP}. The systematic uncertainty in the resultant energy scale in the 5.4-9 keV range has been verified in flight \citep{10.1117/12.3019276} using on-board calibration sources as 0.3 eV. In addition, we use a dedicated calibration pixel that is part of the detector array but placed outside the FOV and is continuously illuminated by a collimated $^{55}$Fe x-ray source. The calibration pixel is used to track the efficacy of the sparse sampling of the energy scale reconstruction function during the observation, outside of the gain monitoring fiducials. For this observation, the energy scale reconstruction was excellent, with a calibration pixel energy scale error of only 0.05 eV at 5.9 keV. As is standard practice for Resolve, we add this in quadrature to the in-flight calibration uncertainty to yield an energy scale uncertainty for this observation of 0.3 eV in the band 5.4-9 keV. Spectra were extracted from the full array for each observation, excluding pixel 27, which has a history of gain jumps that are not well tracked by the Resolve gain-tracking strategy. This is standard practice for Resolve observations. Only high-resolution primary (Hp or ITYPE=0) events were included in the analysis.
 The Redistribution Matrix Function (RMF)
was generated using 
{\tt rslmkrmf} with the
latest CalDB and using all grades excluding 
Ls-events. 
We separated 
the electron loss continuum 
component as a more coarsely 
binned response file.
The anciliary response 
files (ARF) were constructed
with {\tt xaarfgen}, using 
point source mode with 
RA and DEC fixed to
the nominal RA and DEC of the
Observation ID  300002020.

\begin{table*}
\centering
\caption{Best-fit parameters of the continuum + photoionization, continuum+reflection, and continuum + photoionization + reflection models with 90\% uncertainties, derived from fitting the XRISM observation of GX 340+0 within the energy range 3–8 keV.}
\setlength{\tabcolsep}{6pt} 
\renewcommand{\arraystretch}{1.2} 
\begin{tabular}{c|c|c|c|c}
\hline
\textbf{Model Component} & \textbf{Parameter} & \textbf{Photoionization} & \textbf{Reflection} & \textbf{Photoion.+Reflect.} \\
\hline
\texttt{tbabs} &  $N_{\mathrm{H, ISM}}$/ cm$^{-2}$ ($\times 10^{22}$) & $7.52 \pm 0.02$ &  $7.77 \pm 0.02$ & $7.82 \pm 0.03$ \\
\hline
\texttt{diskbb} & $T_{\mathrm{in}}$ (keV) & $1.73 \pm 0.01$ & $1.74 \pm 0.01$ & $1.67 \pm 0.02$ \\
 & norm & $76.91 \pm 0.51$ & 75.97$^{+0.58}_{-0.78}$  & $88.02 ^{+0.65}_{-0.82}$ \\
\hline
{$PIE^{\text{em, narrow (1)}}$} & log $\xi$ & $2.70\pm 0.04$ & — & $2.70 \pm 0.05$ \\
 & log $N_{\mathrm{H, photoionized}}$/ cm$^{-2}$ & $23.22 \pm 0.05$ & — & $23.19 \pm 0.06$ \\
 & \( ^\dagger\rm broadening (\text{km~s$^{-1}$}) \) & $358.20^{+40.22}_{-39.76}$ & — & $357.53^{+47.35}_{-46.34}$ \\
\hline
{$PIE^{\text{em, narrow (2)}}$} & log $\xi$ & $1.40 \pm 0.02$ & — & $1.40 \pm 0.03$ \\
 & log $N_{\mathrm{H, photoionized}}$/ cm$^{-2}$ & $22.50 \pm 0.03$ & — & $22.48 \pm 0.03$ \\
 &  \( ^\dagger\rm broadening (\text{km~s$^{-1}$}) \) &  $358.20^{+40.22}_{-39.76}$ & — & $357.53^{+47.35}_{-46.34}$ \\
\hline
\texttt{vashift (broad emission)} & \( \rm velocity (\text{km~s$^{-1}$})^\# \) & $-2375.56 \pm 127.10$ & — & $-2375.63^{+162.24}_{-146.99}$ \\
{$PIE^{\text{em, broad}}$} & log $\xi$ & $2.83 \pm 0.04$ & — & $3.14 \pm 0.06$ \\
 & log $N_{\mathrm{H, photoionized}}$/ cm$^{-2}$ & $22.95 \pm 0.02$ & — & $22.87 \pm 0.05$ \\
 & \( ^{\dagger\dagger}\rm broadening (\text{km~s$^{-1}$}) \) & $805.97^{+72.38}_{-74.65}$ & — & $804.86^{+88.92}_{-80.03}$ \\
\hline
\texttt{vashift (broad absorption)} & \( ^\ast\rm velocity (\text{km~s$^{-1}$}) \) & $-2375.56 \pm 127.10$ & — & $-2375.63^{+162.24}_{-146.99}$ \\
{$PIE^{\text{abs, broad}}$} & log $\xi$ & $3.65 \pm 0.04$ & — & $4.14 \pm 0.06$ \\
 & log $N_{\mathrm{H, photoionized}}$/ cm$^{-2}$ & $22.95 \pm 0.02$ & — & $22.87 \pm 0.05$ \\
 & \( ^{\dagger\dagger}\rm broadening (\text{km~s$^{-1}$}) \) & $805.97^{+72.38}_{-74.65}$ & — & $804.86^{+88.92}_{-80.03}$ \\
\hline
\texttt{relxillNS}  & Index & — & $2.77 \pm 0.30$ & $3.41^{+0.61}_{-0.53}$ \\
 & $a$ & — & 0$^{f}$ & 0$^{f}$  \\
 & Incl (deg) & — &$33.75^{+0.67}_{-1.04}$ & $40.93^{+1.06}_{-1.75}$  \\
  &  R$_{in}$(R$_{ISCO}$) & — & $1.60^{+0.16}_{-0.23}$& $2.35^{+0.24}_{-0.33}$ \\
  
  &  kT$_{bb}$ (keV) & — & $2.50^{+0.25}_{-0.27}$ & $2.99^{+0.34}_{-0.30}$ \\

 & Ionization (log $\xi$) & — & $2.52^{+0.07}_{-0.06}$ & $2.11 \pm 0.10$ \\
 & A$_{fe}$  & — & $9.9^{+0.1}_{-0.9}$  & $0.84^{+0.09}_{-0.10}$ \\

  &  log n$_{e}$ (cm$^{-3}$)& — & 19$^{f}$ & 19$^{f}$ \\
 
  &  Norm ($\times 10^{-3}$) & — & $1.22 \pm 0.10$ & $2.25 \pm 0.19$ \\
\hline
C-stat/dof & &10622.68/9982 & 10308.22/9989 & 10247.58/9975 \\
\hline
\end{tabular}
$^{\dagger}$-- Parameters tied together during spectral fitting.
$^{\dagger\dagger}$-- Parameters tied together during spectral fitting.
$^{\ast}$-- Parameters tied together during spectral fitting.
$^{f}$-- Parameters frozen  during spectral fitting.
\label{tab:photoionization_model}
\end{table*}

\section{Spectral fitting}\label{sec:3}

{ We performed spectral fitting in \textsc{XSPEC} version 12.14.1 \citep{1996ASPC..101...17A} and modeled the emission and absorption components with $\Cloudy$.
To simulate emission
lines from photoionization equilibrium (PIE) plasma
in $\Cloudy$, we
adopted spectral energy distributions (SED) of GX 340+0
within 2-10 keV
from Resolve observation as there were no simultaneous observations coordinated with other missions during the observation considered here.
We adopted
\citet{2000ApJ...542..914W}
solar abundance
table. 
The metallicity was fixed
to solar value
and electron
density to $10^{14} \, \mathrm{cm}^{-3}$ \citep{2016ApJ...822L..18M}.
}
Spectral fitting was conducted in the 3--8~keV range to focus on prominent Fe and Ca emission and absorption features. The best-fit parameters were determined by minimizing the C-statistic \citep{1979ApJ...228..939C}, and all reported uncertainties correspond to 90\% confidence level.

\begin{figure*}
\centering
\begin{tabular}{cc}
\includegraphics[width=0.5\textwidth]{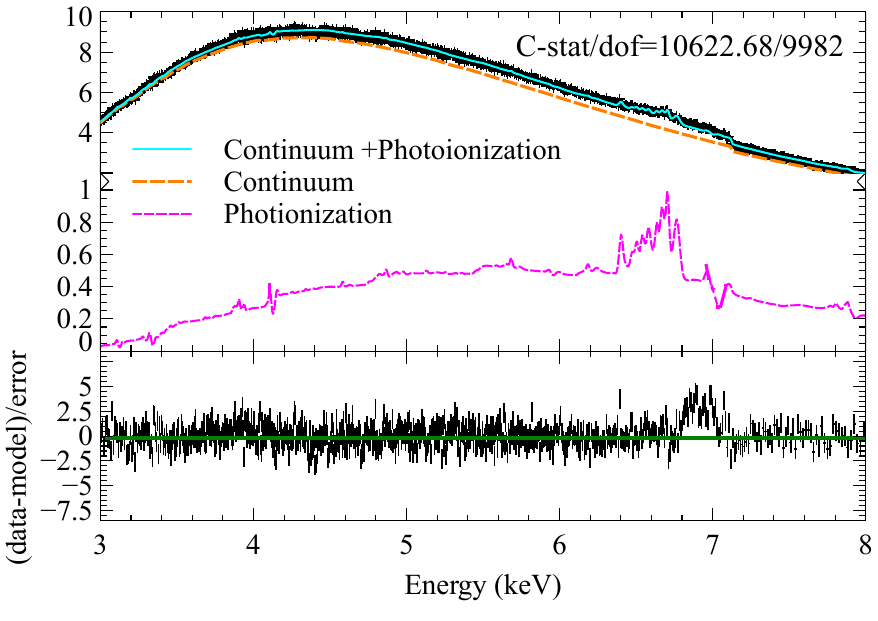} &  \includegraphics[width=0.5\textwidth]{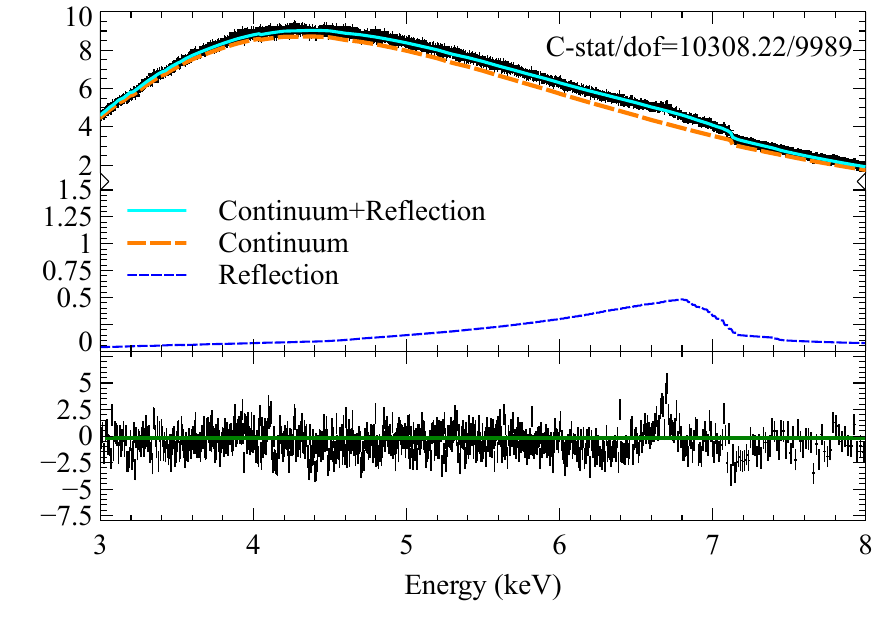}\\
\end{tabular}
\begin{tabular}{c}
\includegraphics[width=0.5\textwidth]{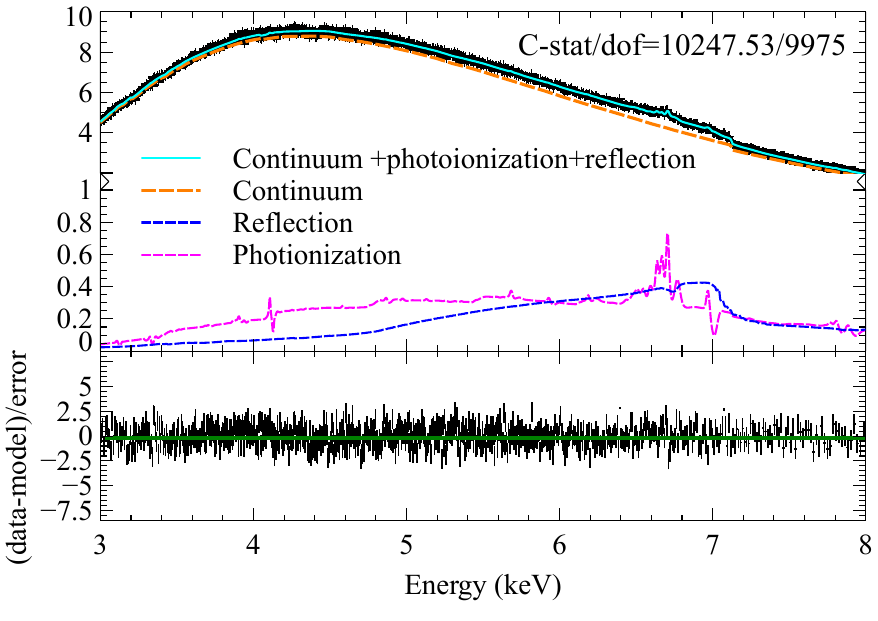}
\end{tabular}
\caption{
XRISM/Resolve spectrum of GX~340+0, shown with various best-fit models (cyan solid line in all panels) across the 3--8~keV energy range. The underlying continuum component is indicated by an orange dashed line in each case. 
{\it Top left panel}: Continuum + photoionization model. The photoionized emission and absorption components, comprising both narrow and broad features, are shown with a magenta dashed line. 
{\it Top right panel}: Continuum + reflection model. The relativistic reflection component is shown with a blue dashed line. 
{\it Bottom panel}: Combined continuum + photoionization + reflection model. The photoionized and reflection components are shown with magenta and blue dashed lines, respectively. Residuals for each fit are shown in the corresponding lower sub-panels.
}
\label{fig:3to8}
\end{figure*}

\subsection{Continuum modeling}

Figure \ref{f:fig1} illustrates the Resolve spectrum in the 3–8 keV range for observation 2. The continuum was modeled using  a multi-temperature blackbody (\texttt{diskbb}) within the \texttt{XSPEC} framework, with interstellar absorption accounted for by the \texttt{tbabs} component  (\texttt{tbabs}$\times$(\texttt{diskbb}); \citealt{1984PASJ...36..741M}).  The \texttt{tbabs} component utilized abundances from \citet{2000ApJ...542..914W} and cross sections from \citet{1995A&AS..109..125V}. This model provides a good representation of the continuum in the 3–8 keV energy range. The best-fit continuum model yields
C--stat/d.o.f of 11546.92/9999, indicating an acceptable fit to the continuum. 
The hydrogen column density was treated as a 
free parameter and constrained to a best-fit 
value of   $N_{\mathrm{H, ISM}}$/ cm$^{-2}$ 
($\times 10^{22}$) = $7.69 \pm 0.07$ (the ISM 
properties in GX 340+0 are analyzed in detail in 
\citet{2025arXiv250608751C}. The temperature of the 
\texttt{diskbb} component was determined to be 
$kT = 1.81 \pm 0.01 \, \mathrm{keV}$, while the 
normalization was estimated to be $65.66 \pm 
0.84$ (in units of
$(R_{\text{in}}/D_{10})^2 \cos\theta$,
where $R_{\text{in}}$ is the inner disk 
radius in km, $D_{10}$ is the distance to
the source in units of 10 kpc, and 
$\theta$ is the disk inclination angle).
The sub-figures at the bottom of Figure \ref{f:fig1} show the residuals of the observed spectra after fitting, calculated as \((\text{data} - \text{model}) / \text{error}\). Residuals are most prominent in the region of the Fe \(\text{XXV}\) He-\(\alpha\) complex between 6.55 and 6.71 keV, with a subtle residual near the Ca \(\text{XX}\) Ly-\(\alpha\) complex between 4.0 and 4.3 keV.

\begin{figure*}
\centering
\begin{minipage}{0.48\textwidth}
    \centering
    \includegraphics[width=\textwidth]{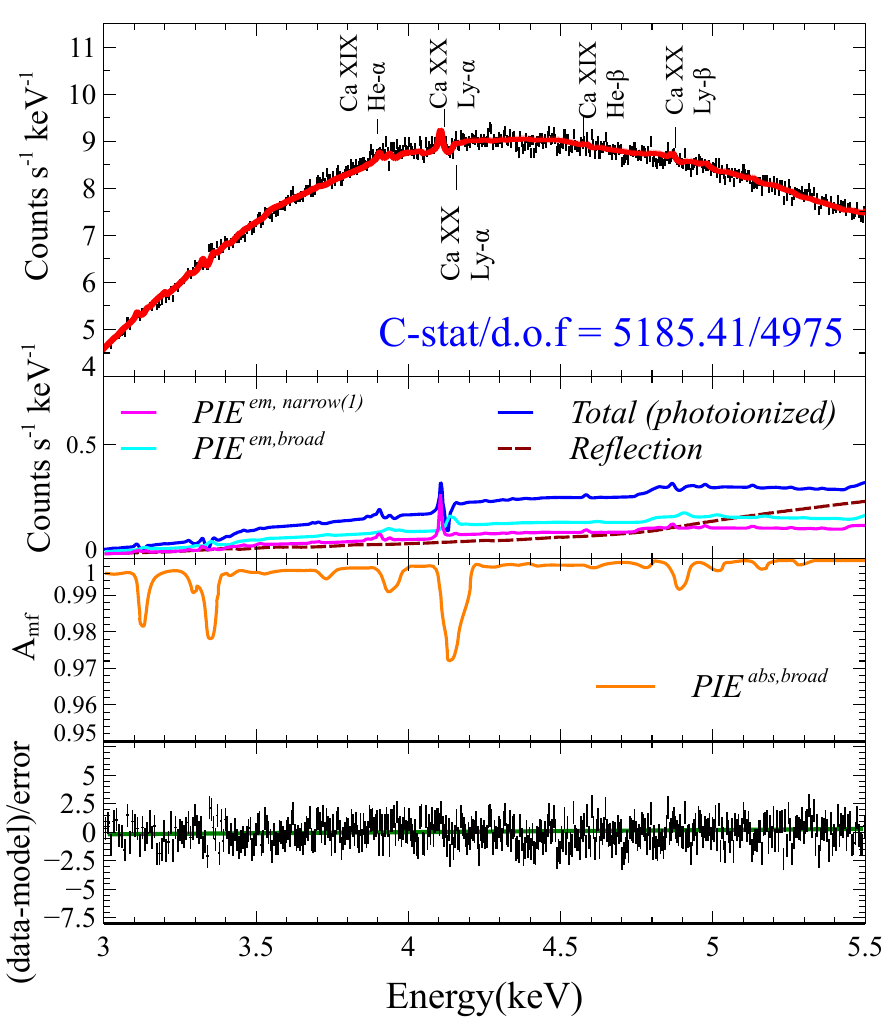}
    \label{fig:a}
\end{minipage}%
\hfill
\begin{minipage}{0.48\textwidth}
    \centering
    \includegraphics[width=\textwidth]{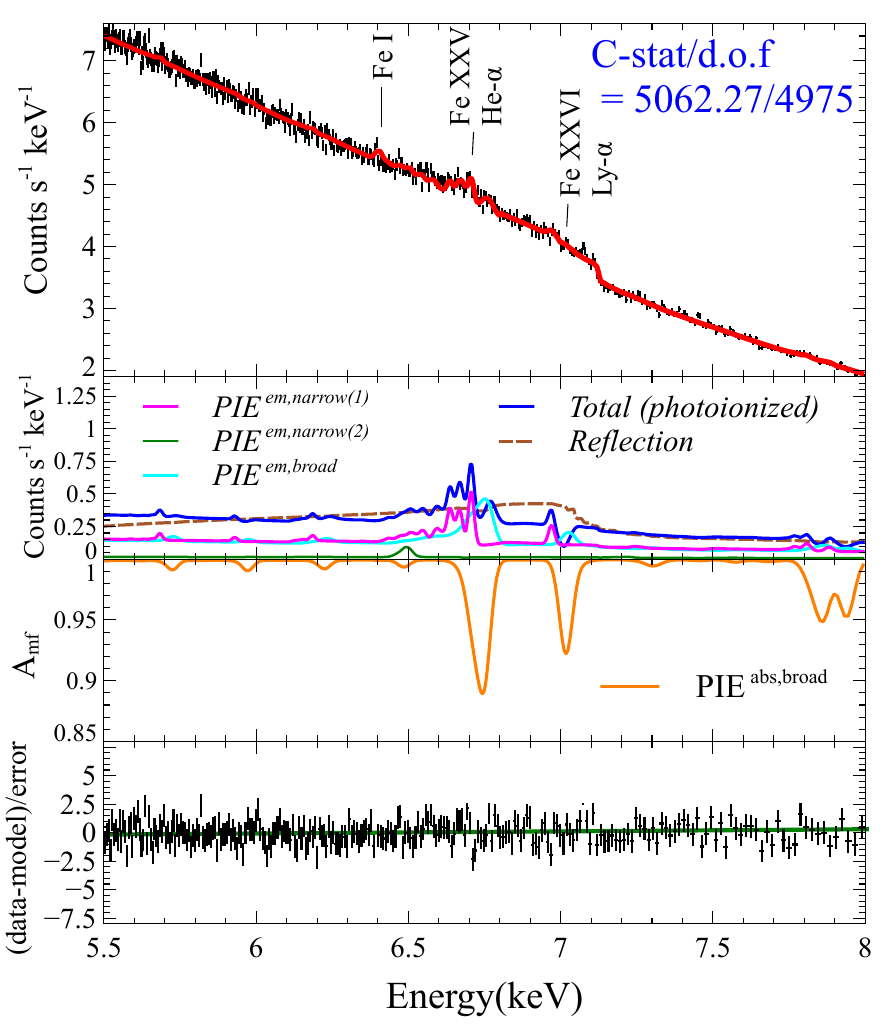}
    \label{fig:b}
\end{minipage}
\vspace{-0.5cm}

\begin{minipage}{0.35\textwidth} 
    \centering
    \includegraphics[width=\textwidth]{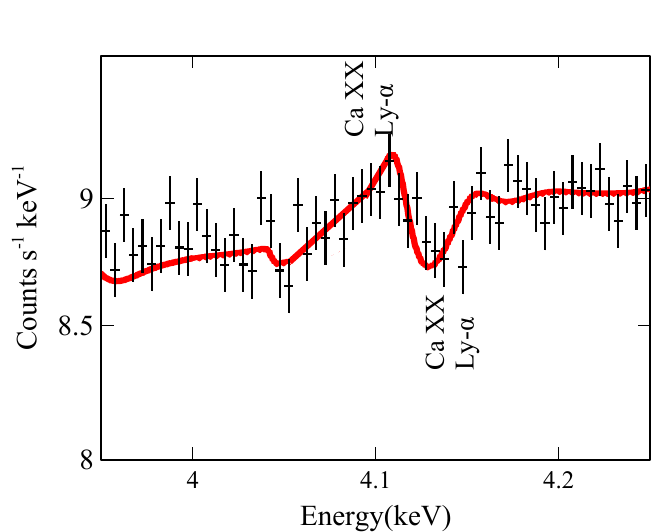}
\end{minipage}
\hspace{0.25cm} 
\begin{minipage}{0.35\textwidth}
    \centering
    \includegraphics[width=\textwidth]{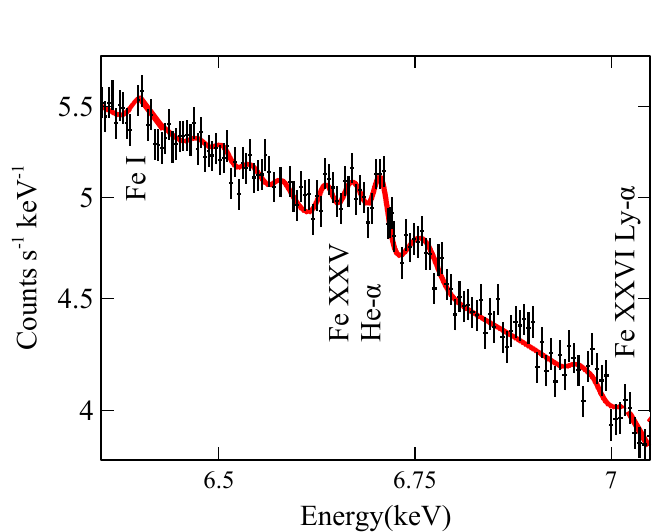}
\end{minipage}
\vspace{0.5cm}

\caption{ 
XRISM/Resolve spectrum of GX~340+0, overlaid with the best-fit continuum, photoionized emission, and reflection models, along with their individual components. The left panel covers the 3--5.5~keV range, highlighting Ca~\textsc{xx} emission and absorption lines. The right panel spans 5.5--8.0~keV, showing Fe~K emission and absorption features.
The second subpanels display the photoionized emission (both narrow and broad), modeled with $\Cloudy$, and the reflection component, modeled with \texttt{relxillNS}, along with the combined photoionized component.
Absorption is modeled as a multiplicative component. The third subpanels show the corresponding dimensionless multiplicative factor, \( A_{\mathrm{mf}} \). The fourth subpanels present the residuals between the observed data and the model.
 Bottom panels:  A magnified view of the Ca XX and Fe K features. }

\label{fig:247fit_2}
\end{figure*}

\begin{figure*}
\centering
\begin{minipage}{0.45\textwidth}
    \centering
    \includegraphics[width=\textwidth]{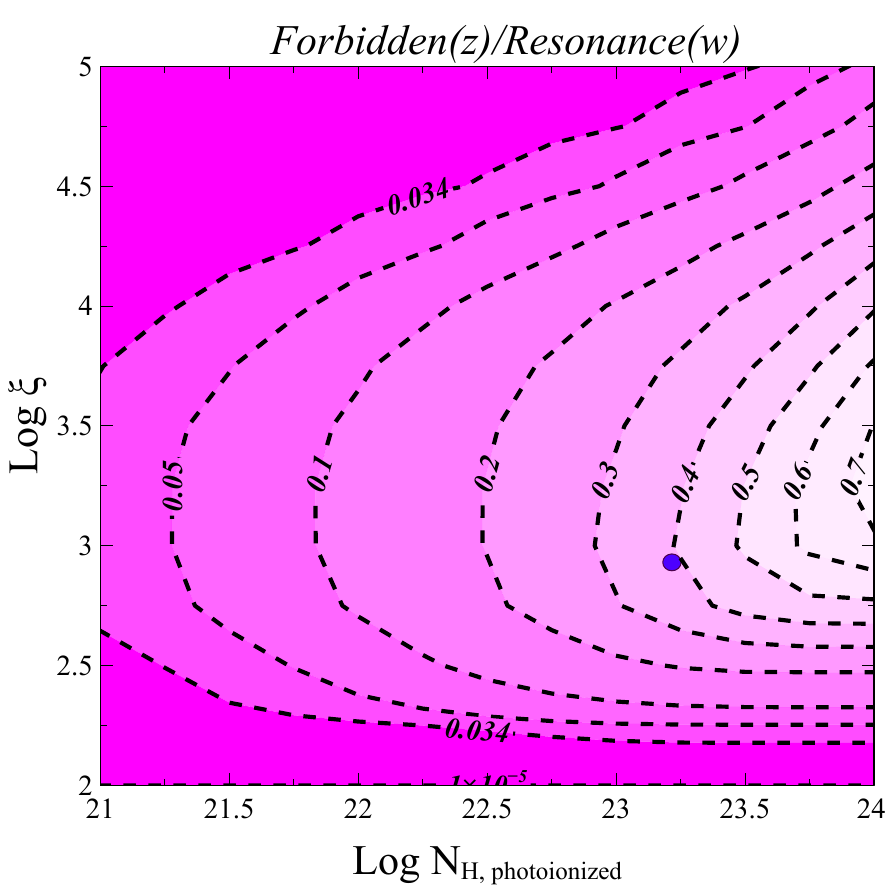}
    \label{fig:a}
\end{minipage}%
\hspace{0.02\textwidth} 
\begin{minipage}{0.45\textwidth}
    \centering
    \includegraphics[width=\textwidth]{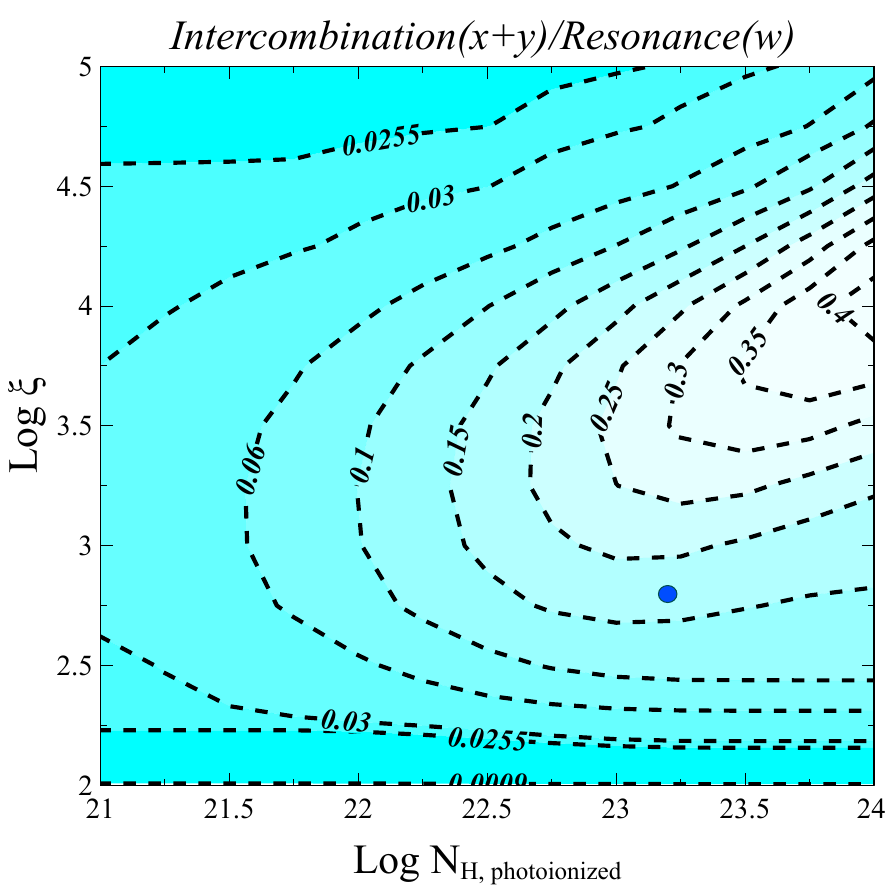}
    \label{fig:b}
\end{minipage}

\begin{minipage}{0.95\textwidth}
    \centering
    \includegraphics[width=\textwidth]{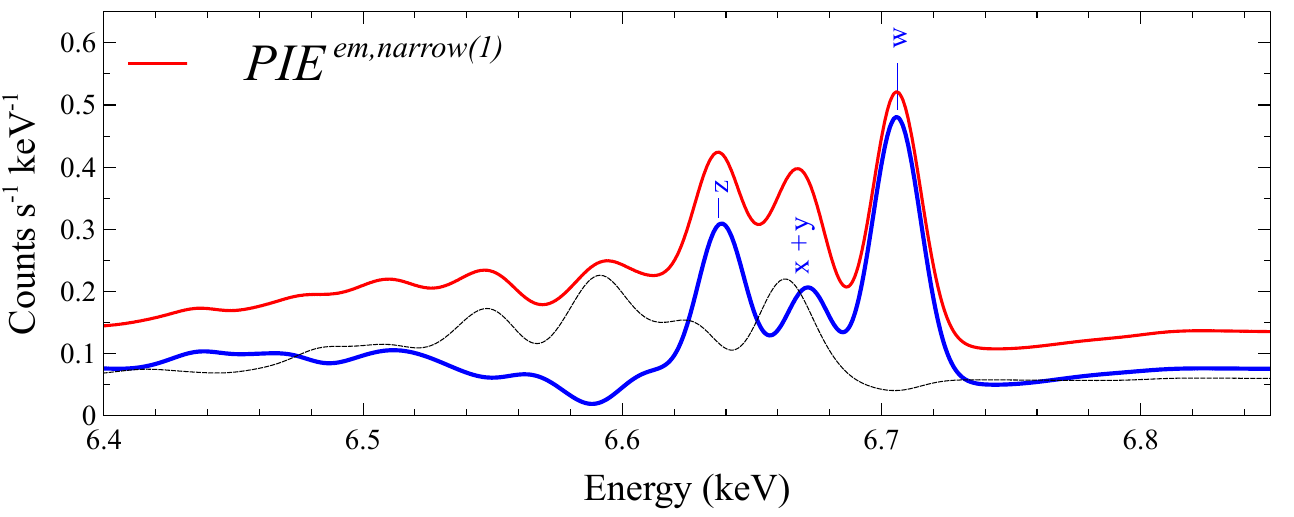} 
    \label{fig:c}
\end{minipage}

\caption{ 
{Top Panels:} Variation of \(z/w\) and \((x+y)/w\) ratios as a function of \(\log \xi\) and \(\log N_{\text{H, photoionized}}\) for PIE plasma, calculated using $\Cloudy$ for an electron density \(n_e = 10^{14} \, \text{cm}^{-3}\). The best-fit parameters within the 3--8 keV range for GX 340+0 are marked with blue circles for $PIE^{\text{em, narrow(1)}}$.
{Bottom Panel:} Best-fit rest-frame narrow photoionized emission component ($PIE^{\text{em, narrow(1)}}$) from Figure 2 (shown in red). The Fe~\textsc{xxv} He$\alpha$ lines are displayed in blue, where \( z \) and \( w \) are resolved, while \( x \) and \( y \) are blended. The sum of the emission from Fe~\textsc{xxiv}, Fe~\textsc{xxiii}, and Cr~\textsc{xxiii} lines is shown with a black dashed line.
}
\label{fig:CaseCD}
\end{figure*}

\subsection{Continuum + Photoionization modeling}\label{conpho}

$\Cloudy$ determines the line emission/absorption by solving a fully coupled system of equations giving radiative transitions between levels and charges, as originally described in Section III of \citet{1989ApJ...347..640R}. 
It also incorporates detailed radiative transfer and microphysical processes, including continuum pumping, scattering, absorption (such as electron scattering escape), and Resonance Auger Destruction \citep{2020RNAAS...4..184C, 2020ApJ...901...68C, 2020ApJ...901...69C, 2021ApJ...912...26C, 2022ApJ...935...70C}. Refer to Section \ref{physics} for a detailed discussion of these physical processes and their role in shaping the observed spectrum.

For simultaneous modeling of both emission and absorption features, we constructed theoretical grids based on two key parameters: the ionization parameter ($\xi$) and the hydrogen column density of the photoionized plasma ($N_{\rm H, photoionized}$). A logarithmic grid was generated for $\xi$, spanning values of $\log \xi$ from 0.5 to 5 in increments of 0.1 dex. Similarly, a logarithmic grid was constructed for $N_{\rm H, photoionized}$, covering $\log N_{\rm H, photoionized}$ values from 20 to 24, also with 0.1 dex increments. This resulted in a extensive set of 2,000 distinct grid points for our analysis. The output files from these calculations were converted into FITS format using the $\Cloudy$--{\tt XSPEC} interface \citep{2006PASP..118..920P} to ensure compatibility with \texttt{XSPEC}. Within the $\Cloudy$ model, turbulent and bulk velocities were set to zero. Velocity broadening was applied externally and treated as a free parameter.

We began the analysis with a single emission or absorption component, incrementally incorporating additional components as needed to achieve an optimal fit. Emission line components were represented using an \texttt{atable} (additive tabular model format), while absorption was applied to the combined continuum and emission spectrum using the \texttt{mtable} (multiplicative tabular model) format, as described in \citet{2006PASP..118..920P}. The final model comprised three emission components and one absorption component (see equation 1). Among the emission components, two were narrow with minimal velocity broadening in the rest frame, while one was broad and blueshifted. The absorption component was also broad and exhibited the same blueshift as the broad emission component. We refer to the narrow emission components as \( \text{PIE}^{\text{em, narrow(1)}} \) and \( \text{PIE}^{\text{em, narrow(2)}} \), the broad emission component as \( \text{PIE}^{\text{em, broad}} \), and the broad absorption component as \( \text{PIE}^{\text{abs, broad}} \).
                
An Fe XXV He\(\alpha\) emission complex was identified in the rest frame. Velocity broadening of \(\sim 360 \,  \text{km~s$^{-1}$}\) in the Fe XXV 
He\(\alpha\) complex caused the intercombination lines (\(x\) and \(y\)) to blend, while the resonance (\(w\)) and forbidden (\(z\))
lines remained resolved. This corresponds to the first \( \text{PIE}^{\text{em, narrow (1)}} \), characterized as a moderately 
ionized plasma with an ionization
parameter of log $\xi$ $\sim$ 2.70 and a high hydrogen column density of 
log N$_{\rm H, photoionized}  \sim \ 23.22$,
which significantly improved the fit by \( \Delta\text{C-stat} = 494.56 \) for 4 additional degrees of freedom. This component successfully 
modeled the rest-frame He\(\alpha\) triplet lines within the Fe XXV complex. 
The same \(\text{PIE}^{\text{em, narrow(1)}}\) component was used to model the rest-frame Ca XX emission, which exhibited a velocity broadening of \(\sim 360 \,  \text{km~s$^{-1}$}\) but was not resolved into the Ly\(\alpha\) doublets. A faint emission from Ca XIX was detected and modeled using the same best-fit parameters. This feature was not resolved into its He\(\alpha\) quadruplet structure.

For modeling the fluorescent line at 6.4 keV, we introduced a lower-ionization component (log $\xi$ $\sim$ 1.40) with   \(\log N_{\rm H, photoionized} \sim 22.50\). 
This corresponds to the second \( \text{PIE}^{\text{em, narrow(2)}} \), which was well-fitted with the same velocity broadening of \(\sim 360 \,  \text{km~s$^{-1}$}\) as the first \( \text{PIE}^{\text{em, narrow (1)}} \) and further improved the fit by  \( \Delta\text{C-stat} = 37.48\) for 3 additional degrees of freedom.\\
An additional Fe XXV He\(\alpha\) emission component, blueshifted by \(\sim 2734 \,  \text{km~s$^{-1}$}\), was introduced to account for a broader emission feature near the Fe K complex. This component corresponds to the only \( \text{PIE}^{\text{em, broad}} \), yielding a significant improvement in the fit with \( \Delta\text{C-stat} = 287.44 \) for 5 extra degrees of freedom.
 The broadening of this feature was determined to be twice as large as that of the emission components used to model the Fe XXV He\(\alpha\) complex in the rest frame, \(\sim 800 \,  \text{km~s$^{-1}$}\). The ionization parameter was estimated to be  log $\xi$ $\sim$ 2.83, with a hydrogen column density of 
log N$_{\rm H, photoionized}  \sim \ 22.95$.\\
An absorption line associated with the Ca XX Ly\(\alpha\) transition was identified, showing a blueshift of \(\sim 2735 \,  \text{km~s$^{-1}$}\), corresponding to the only \( \text{PIE}^{\text{abs, broad}} \). This line exhibited velocity broadening of \(\sim 800 \,  \text{km~s$^{-1}$}\), which is identical to that observed in \( \text{PIE}^{\text{em, broad}} \). The broadening prevented the resolution of the expected doublet structure. The absorption feature was modeled using a photoionized component characterized by an ionization parameter of \(\log \xi \sim 3.65\) and a hydrogen column density of \(\log N_{\rm H, photoionized} \sim 22.95\), resulting in a fit improvement of \( \Delta\text{C-stat} = 104.76 \) for 2 extra degrees of freedom. The final best-fit model, derived from the simultaneous incorporation of all previously identified photoionized emission and absorption components, yields a C-statistic of 10622.68 for 9982 degrees of freedom. 

The complete XSPEC model can be described as:
\begin{dmath}
M_{1}= \texttt{tbabs} \times \Big( \texttt{diskbb} + \text{PIE}^{\text{em, narrow(1)}} + \text{PIE}^{\text{em, narrow(2)}} \\ + \texttt{vashift} \times \text{PIE}^{\text{em, broad}} \Big) \times \texttt{vashift} \times \text{PIE}^{\text{abs, broad}}
\end{dmath}
The complete list of best-fit parameters for the continuum combined with all photoionized components is presented in Table~\ref{tab:photoionization_model}. The close agreement in velocity broadening, blueshift, and column density between \( \text{PIE}^{\text{abs, broad}} \) and \( \text{PIE}^{\text{em, broad}} \) strongly suggest a common physical origin, most likely within an outflowing wind.
The top panel of Figure~\ref{fig:3to8} displays the observed spectrum with the full model overlaid, incorporating both the continuum and photoionized components. A prominent broad residual visible in the lower sub-panel indicates potential presence of a relativistic reflection component, which is further investigated in Sections \ref{conref}  and \ref{c+p+r}.

\subsection{Continuum + Reflection modeling}\label{conref}

To assess the contribution of relativistic disk reflection, we incorporated the relxillNS model \citep{2013ApJ...768..146G, 2022ApJ...926...13G, 2014MNRAS.444L.100D}, which self-consistently models reflection from the accretion disk assuming irradiation by a neutron star.  Given the independent treatment of the illuminating continuum, we fixed Refl$_{frac}$ = $-1$. The spectral analysis was carried out with the emissivity indices described by a single free parameter $q$ (with $q_1 = q_2 = q$), and the outer disc radius fixed at $R_{\mathrm{out}} = 990  R_{\mathrm{g}}$, where $R_{\mathrm{g}}$ is the Schwarzschild radius.
The dimensionless spin parameter was set to $a = 0$ \citep{2008ApJS..179..360G, 2011ApJ...731L...7M}. 
We fixed a disk electron density of \( n_e = 10^{19} \,\mathrm{cm}^{-3} \), which is the maximum value currently supported in \texttt{relxillNS}.
The inner disc radius ($R_{\mathrm{in}}$), emissivity index ($q$) , inclination angle ($\theta$),  the temperature of the blackbody incident spectrum within \texttt{relxillNS} (\( kT_{\mathrm{bb}} \)), ionization parameter ($\xi$), iron abundance ($A_{\mathrm{Fe}}$), and the normalization were all allowed to vary freely.

 The full spectral model adopted in XSPEC is:
\begin{equation}
M_{2} = \texttt{tbabs} \times (\texttt{diskbb} + \texttt{relxillNS})
\end{equation}
 The best-fit parameters for the relativistic reflection component are:  $q \sim 2.77$, $R_{\mathrm{in}} \sim 1.60$ (in units of $R_{\mathrm{ISCO}}$, the radius of the innermost stable circular orbit),  $i \sim 33.75^\circ$, log $\xi \sim 2.52$, \( kT_{\mathrm{bb}} \) $\sim$ 2.50 keV,  $A_{\mathrm{Fe}} \sim 9.9$ (relative to solar), and model normalisation $\sim 1.22 \times 10^{-3}$. 
 
 The supersolar iron abundance, which approaches the upper limit of the model, is not unusual in reflection modeling and may arise from a combination of physical effects and modeling limitations. The likely contributor is the reflection component compensating for excess flux associated with high-ionization line emission from photoionized gas by artificially inflating $A_{\mathrm{Fe}}$.
 As demonstrated in Section \ref{c+p+r}, introducing a physically motivated photoionized emission component yields an improved fit and removes the need for an anomalously high iron abundance, indicating that the extreme $A_{\mathrm{Fe}}$ is more likely a modeling artifact rather than a true abundance enhancement.\footnote{Alternative explanations for supersolar iron abundances have been proposed in the literature.  A supersolar iron abundance may also result from reflection models compensating for unresolved high-density effects in the inner disk  \citep{2018ApJ...855....3T}, where electron densities are expected to exceed $n_{\mathrm{e}} > 10^{20}$ cm$^{-3}$, well above the $n_{\mathrm{e}} \leq 10^{19}$ cm$^{-3}$ limit assumed in \texttt{relxillNS}, as also discussed in \citet{2025arXiv250706289L}. Additionally, radiative levitation has been proposed as a mechanism by which heavy elements can accumulate near the disk surface through radiation pressure, potentially leading to enhanced apparent abundances in the reflection spectrum \citep[e.g.,][]{2012ApJ...755...88R}.}

The full set of best-fitting parameters is reported in Table~\ref{tab:photoionization_model}. The fit yields a C-statistic of 10308.22 for 9989 degrees of freedom, a statistically acceptable fit. Nonetheless, the residuals shown 
in the lower sub-panel of the middle panel of Figure \ref{fig:3to8} exhibits significant narrow emission and absorption features not fully captured by a reflection-only model. 
Additionally, the elevated iron abundance further indicates that a reflection-only model does not
adequately capture the full complexity of the spectral features.
This motivates the addition of a photoionization component alongside the reflection model, as examined in the following section, to achieve a more complete fit to the spectra with minimal residuals.

\subsection{Continuum+ Photoionization+ Reflection}\label{c+p+r}

As both the best-fit continuum + photoionization model (Section~\ref{conpho}) and the best-fit continuum + reflection model (Section~\ref{conref}) exhibited noticeable residuals, we adopted a combined model incorporating both photoionized emission and relativistic reflection components to achieve a more comprehensive description of the spectrum.
 
The complete model implemented in \textsc{XSPEC} was:
\begin{dmath}
M_{3} = \texttt{tbabs} \times \Big( \texttt{diskbb} + \texttt{relxillNS} + \text{PIE}^{\text{em, narrow(1)}} + \text{PIE}^{\text{em, narrow(2)}} \\
+ \texttt{vashift} \times \text{PIE}^{\text{em, broad}} \Big) \times \texttt{vashift} \times \text{PIE}^{\text{abs, broad}}
\end{dmath}

The treatment of parameters, whether fixed, free, or tied, was identical to that in the individual models described in Sections~\ref{conpho} and~\ref{conref}.
Table ~\ref{tab:photoionization_model} lists the best-fit parameters obtained from modeling the continuum together with the photoionized and reflection components.  The first photoionized emission component, \( \text{PIE}^{\text{em, narrow(1)}} \), shows a slight adjustment in its parameters, with log\,\(\xi \sim 2.70\) and a hydrogen column density of log\,\(N_{\rm H,\,photoionized} \sim 23.19\). The second component, \( \text{PIE}^{\text{em, narrow(2)}} \), retains the same ionization parameter of log\,\(\xi \sim 1.40\), with a slightly reduced column density of log\,\(N_{\rm H,\,photoionized} \sim 22.48\). Both components were well fitted with a velocity broadening of \(\sim 360\,\text{km~s}^{-1}\), which remained unchanged from the earlier fit described in Section \ref{conpho}.

The parameters of \( \text{PIE}^{\text{em, broad}} \), blueshifted by \(\sim 2735\,\text{km\,s}^{-1}\), show a moderate change, with the ionization parameter increasing to log\,\(\xi \sim 3.14\) and the hydrogen column density decreasing to log\,\(N_{\rm H,\,photoionized} \sim 22.87\) compared to the continuum+photoionization modeling scenario. The corresponding absorption component, \( \text{PIE}^{\text{abs, broad}} \), exhibits the same blueshift and shows a further increase in ionization, with log\,\(\xi \sim 4.14\), while maintaining a similar column density of log\,\(N_{\rm H,\,photoionized} \sim 22.87\).
The updated best-fit parameters for the relativistic reflection component are as follows:  \( q \sim 3.41 \),  \( R_{\mathrm{in}} \sim 2.35 \, R_{\mathrm{ISCO}} \),  \( i \sim 40.94^\circ \), \(kT_{\mathrm{BB}} \sim 2.99 \, \text{keV} \),  log\,\(\xi \sim 2.10\), \( A_{\mathrm{Fe}} \sim 0.84 \) (relative to solar), and normalization \( \sim 2.25 \times 10^{-3} \).
Notably, when the model includes both photoionization and reflection components, the inferred iron abundance becomes subsolar, in contrast to the supersolar values obtained in Section \ref{conref},  where the reflection component may have compensated for photoionized features not explicitly modeled.

The bottom panel of Figure \ref{fig:3to8} shows the observed spectrum with $M_{3}$ overlaid, capturing the combined contributions  of photoionization and reflection.  The fit with $M_{3}$ yields a C-statistic of 10247.58 for 9975 degrees of freedom. Compared to $M_{1}$, this corresponds to an improvement of $\Delta\mathrm{C\text{-}stat} = 379.3$ with 7 additional degrees of freedom, and an improvement of $\Delta\mathrm{C\text{-}stat} = 60.64$ over $M_{2}$, with 12 additional degrees of freedom.
The incorporation of both photoionized and relativistic reflection components in $M_{3}$ effectively eliminates the residual features observed in fits with $M_{1}$ and $M_{2}$ (see lower sub-panels in Figure~\ref{fig:3to8}), suggesting that a physically consistent description of the spectrum requires the inclusion of both photoionized emission and relativistic reflection. The top panel of Figure \ref{fig:247fit_2} displays a focused view of the 3.0–5.5 keV and 5.5–8.0 keV energy ranges (shown separately for clarity in presentation) along with individual narrow and broad photoionized emission and absorption features. The bottom panel presents a magnified view of the Ca XX and Fe K regions.

\section{Physics of the photoionized lines: $\fexxv$ He$\alpha$ complex}\label{physics}

In a stationary photoionized plasma, forbidden (\(z\)) and intercombination (\(x, y\)) lines are typically stronger than the resonance (\(w\)) line at high column densities (\(\log N_{\text{H}} > 23\); \citealt{2004A&A...414..979C}). In our observations, however, this trend is reversed: for both the broad and narrow emission components (\( \text{PIE}^{\text{em, narrow(1)}} \) and \( \text{PIE}^{\text{em, broad}} \)), the resonance (\(w\)) line is stronger than the forbidden (\(z\)) and intercombination (\(x, y\)) lines, as illustrated in the middle and bottom panels of Figure \ref{fig:CaseCD}.
To investigate this apparent discrepancy, we analyze the behavior of the emission lines in the Fe XXV He-$\alpha$ complex. The intensities of photoionized emission lines can be either enhanced or suppressed, depending on the interplay of two primary mechanisms:

\begin{itemize}
    \item 
    
Radiative excitation of atoms or ions by continuum photons in the spectral energy distribution (SED) can enhance line intensities. This process, known as continuum pumping, preferentially boosts line intensities with large oscillator strengths. Within the Fe XXV He-$\alpha$ complex, the resonance ($w$) line is most strongly affected by this mechanism \citep{2021ApJ...912...26C}.

\item  Conversely, line intensities can be suppressed due to photoelectric absorption and electron scattering, which become increasingly significant as optical depth increases \citep{2022ApJ...935...70C}.

\end{itemize}
\citet{2021ApJ...912...26C} demonstrated that resonance line intensities can increase by a factor of up to \(\sim 30\) due to continuum pumping. However, this enhancement decreases as column density increases, since optical depth, which is proportional to column density \citep{1986rpa..book.....R}, becomes significant. When the photoionized gas is optically thin, the line formation process is categorized as Case C, whereas optically thick conditions correspond to Case D. Thus, line formation is classified as Case C when continuum pumping dominates, and transitions to Case D when optical depth effects dominate. The interplay between enhancement via continuum pumping and suppression due to optical depth is governed by parameters such as the column density and the ionization parameter \citep{2022ApJ...935...70C}. These factors ultimately determine the relative intensity of the resonance (\(w\)) line compared to the forbidden ($z$) and intercombination ($x,y$) lines in both the narrow and broad photoionization emission (\(\text{PIE}^{\text{em, narrow(1)}}\) and \(\text{PIE}^{\text{em, broad}}\))   observed in GX 340+0.\\
To visualize  how the column density and the ionization parameter impact relative line intensities, we have plotted the line ratios \(z/w\) and \((x+y)/w\) in a two-dimensional parameter space, with the ionization parameter (\(\xi\)) on the y-axis and the hydrogen column density (\(N_{\text{H}}\)) on the x-axis, as illustrated in the upper panels of Figure \ref{fig:CaseCD}. At low column densities (\(\log N_{\text{H}} = 21\)), \(z/w\) and \((x+y)/w\) are approximately 0.03, with the resonance (\(w\)) line being \(\sim 30\) times stronger than the forbidden (\(z\)) and intercombination (\(x+y\)) lines. This behavior reflects the dominance of continuum pumping, consistent with the Case C limit. As the column density increases, optical depth effects become more significant, leading to a gradual transition from Case C to Case D conditions. 
By \(\log N_{\text{H}} = 24\), \(z/w\) increases to approximately 0.7, and \((x+y)/w\) rises to around 0.4, marking a clear progression toward the Case D regime, where optical depth effects play a substantial role in shaping the relative line intensities. Although optical depth effects significantly diminish the relative strength of the \(w\) line compared to the \(z\) and \((x+y)\) lines, the \(w\) line remains the dominant feature, with its intensity exceeding the  contributions of \(z\) and \((x+y)\).

\begin{table*}
\centering
\caption{List of photoionized rest-frame spectral lines detected in GX 340+0, including their transitions and energies, modeled using $\Cloudy$. Lines marked with \( \dagger \) are detected but are not individually resolved. 
}
\setlength{\tabcolsep}{10pt} 
\renewcommand{\arraystretch}{1.2} 
\begin{tabular}{ccccccc}
\hline
\textbf{Ion} & \textbf{Energy(keV)} & \textbf{Label} & \textbf{Transition} \\
\hline

Ca XIX$^\dagger$ & 3.86113 & He $\alpha$, z   & $1s^{2}$ ($^1S_0$) – $1s.2p (^3S_1$)\\
Ca XIX$^\dagger$ & 3.88331 & He $\alpha$, y   & $1s^{2}$ ($^1S_0$) – $1s.2p (^3P_1$)\\
Ca XIX$^\dagger$  & 3.88770 &  He $\alpha$, x   & $1s^{2}$ ($^1S_0$) – $1s.2p (^3P_2$)\\
Ca XIX$^\dagger$ & 3.90226 &  He $\alpha$, w   & 1s$^{2}$ ($^1S_0$) – $1s.2p (^1P_1$)\\
Ca XX$^\dagger$  & 4.10014 &   Ly $\alpha_{2}$     & \(1s \, ({}^2S_{1/2}) -2p \, ({}^2P_{1/2})\) \\
Ca XX$^\dagger$ & 4.10749 &   Ly $\alpha_{1}$   &   \(1s \, ({}^2S_{1/2}) -2p \, ({}^2P_{3/2})\)\\
Ca XIX $^\dagger$ & 4.58280 &   He $\beta$   &   1s$^{2}$ ($^1S_0$) – $1s.3p (^1P_1$)\\
Ca XX$^\dagger$  & 4.86192 &   Ly $\beta_{2}$     & \(1s \, ({}^2S_{1/2}) -3p \, ({}^2P_{1/2})\) \\
Ca XX$^\dagger$ & 4.86409 &   Ly $\beta_{1}$   &   \(1s \, ({}^2S_{1/2}) -3p \, ({}^2P_{3/2})\)\\
Fe fluorescence & 6.40000 & K$\alpha$   & - \\
Fe XXIII$^\dagger$ & 6.62911 &  $\beta$   & $1s^{2}2s^{2}$ ($^1S_{0}$) – $1s.2s^{2}.2p (^1P_{1}$)\\
Fe XXV & 6.63659 & He $\alpha$, z   & $1s^{2}$ ($^1S_0$) – $1s.2p (^3S_1$)\\
Fe XXV$^\dagger$ & 6.66757 & He $\alpha$, y   & $1s^{2}$ ($^1S_0$) – $1s.2p (^3P_1$)\\
Fe XXIV$^\dagger$ & 6.65330 &  r  & $1s^{2}2s$ ($^2S_{1/2}$) – $1s.2s.2p (^2P_{1/2}$)\\
Fe XXIV$^\dagger$ & 6.66188 &  q   & $1s^{2}2s$ ($^2S_{1/2}$) – $1s.2s.2p (^2P_{3/2}$)\\
Fe XXIV$^\dagger$ & 6.67658 &  t   & $1s^{2}2s$ ($^2S_{1/2}$) – $1s.2s.2p (^2P_{1/2}$)\\
Fe XXIV$^\dagger$ & 6.67910 &  s   & $1s^{2}2s$ ($^2S_{1/2}$) – $1s.2s.2p (^2P_{3/2}$)\\
Cr XXIII$^\dagger$ & 6.68076 &  -   & $1s^{2}$ ($^1S_0$) – $1s.2p (^1P_1$)\\
Fe XXV$^\dagger$ & 6.68231 &  He $\alpha$, x   & $1s^{2}$ ($^1S_0$) – $1s.2p (^3P_2$)\\
Fe XXV & 6.70040 &  He $\alpha$, w   & 1s$^{2}$ ($^1S_0$) – $1s.2p (^1P_1$)\\
Fe XXVI$^\dagger$  & 6.95197 & Ly $\alpha_{2}$   & \(1s \, ({}^2S_{1/2}) -2p \, ({}^2P_{1/2})\)\\

Fe XXVI$^\dagger$  & 6.97316 & Ly $\alpha_{1}$  & \(1s \, ({}^2S_{1/2}) -2p \, ({}^2P_{3/2})\)\\

\hline

\end{tabular}
\label{tab:model_lines}
\end{table*}
For GX 340+0, the narrow photoionized emission component (\(\text{PIE}^{\text{em, narrow(1)}}\)), as reported in Table 1, yields best-fit parameters of \(\log \xi \sim 2.70\) and \(\log N_{\text{H}} \sim 23.19\). These values, when mapped onto the contour plot in the top panel of Figure \ref{fig:CaseCD},  correspond to the observed line ratios of \(z/w \sim 0.4\) and \((x+y)/w \sim 0.18\), as also displayed in the bottom panel of Figure \ref{fig:CaseCD}.
This demonstrate that the combined effects of ionization parameter and column density govern the observed line properties. Even in a high-column density regime (\(\log N_{\text{H}} > 23\)) in a photoionized plasma, the resonance line (\(w\)) remains dominant, with its intensity surpassing that of both the intercombination (\(x, y\)) and forbidden (\(z\)) lines.


\section{Discussion}

In this work, we present a detailed analysis of the first \textit{XRISM} observation of GX~340+0, obtained during the mission’s performance verification phase. The high-resolution spectrum reveals a rich array of spectral features exhibiting a range of line widths, velocities, and ionization states.

We detect both rest-frame emission lines---such as those from Fe~\textsc{xxv} and Ca~\textsc{xx} as well as blueshifted emission and absorption features from the same ions, indicating the presence of both static and outflowing plasma components. These features are interpreted through detailed photoionization modeling using $\Cloudy$, which constrains the physical properties of the emitting and absorbing gas. In addition, we identify a broad reflection component in the spectrum, modeled with the relativistic reflection framework \texttt{relxillNS}. The full spectral modeling combines both photoionized emission and absorption, along with relativistic reflection, to provide a self-consistent interpretation of the complex X-ray spectrum.

Key findings include the identification of distinct photoionized components, both narrow and broad, which are characterized by 
their ionization parameters, column densities, and velocity broadenings. The narrow
emission component 
(\(PIE^{\text{em, narrow(1)}}\)) 
successfully modeled the rest-frame
Fe XXV and Ca XX features, revealing 
moderately ionized plasma 
(log $\xi$ $\sim$ 2.70) and high column
density (log $N_{\rm H, photoionized}$ $\sim$ 23.19).
The fluorescent line at 6.4 keV
was modeled using a lower-ionization component (log $\xi$ $\sim$ 1.40)
, log $N_{\rm H, photoionized}$ $\sim$ 22.48),
corresponding to \(PIE^{\text{em, narrow(2)}}\),
which shares the same velocity broadening
of \(\sim 360 \,  \text{km~s$^{-1}$}\) as \(PIE^{\text{em, narrow(1)}}\).
Meanwhile, broad emission (\(PIE^{\text{em, broad}}\))
and absorption (\(PIE^{\text{abs, broad}}\)) components were identified,
both exhibiting a shared blueshift of \(\sim 2735 \,  \text{km~s$^{-1}$}\) 
and velocity broadening of \(\sim 800 \,  \text{km~s$^{-1}$}\). 
These broad components exhibit distinct ionization states (log 
$\xi$ $\sim$ 3.14
and log $\xi$ $\sim$ 4.14, respectively)
and significant column density (log $N_{\rm H, photoionized}$ $\sim$ 22.87), 
indicative of a dense region with a moderately ionized outflow region/disk wind.

We estimate the wind launching radius (\(R_L\)) of the disk wind using the following equation \citep{2018ApJ...861...26A}


\begin{equation}
R_L = \sqrt{\frac{L}{n \xi}}
\end{equation}

One of the key challenges in disk wind observations lies in accurately constraining the plasma density. This uncertainty not only affects the characterization of the wind properties but also complicates the determination of the wind launching radius.  For GX 340+0, an independent measurement of the electron density is not available in the literature. Therefore, we adopt the electron density of \(n_e = 10^{14} \, \text{cm}^{-3}\) used in previous studies of GX 340+0 for consistency \citep{2016ApJ...822L..18M}, informed by earlier direct measurements of electron density in stellar-mass black hole systems displaying ionized wind in the Fe K band \citep{2015ApJ...814...87M}.
Based on the continuum, the 2-10 keV luminosity was estimated to be \(L = 9.64 \times 10^{37} \, \text{erg s}^{-1}\), assuming a distance of 11 kpc \citep{2000MNRAS.317....1F}.
With an average ionization parameter of $\log \xi \sim 3.64$ (calculated from $\log \xi \sim 4.14$ for $PIE^{\text{abs, broad}}$ and $\log \xi \sim 3.14$ for $PIE^{\text{em, broad}}$), the corresponding launching radius is estimated to be $\sim$ $1.5 \times 10^{10} \, \text{cm}$. 
A disk wind has been potentially detected in GX 340+0 by \citet{2016ApJ...822L..18M}, who reported a high-velocity wind (\(\sim 0.04c\)). The lower wind velocity ($\sim$ 0.01c) detected in the \textit{XRISM} observation may reflect differences in the system's physical conditions at different epochs or suggest variability in the wind properties over time.

\begin{figure}
\centering
\includegraphics[width=0.45\textwidth]{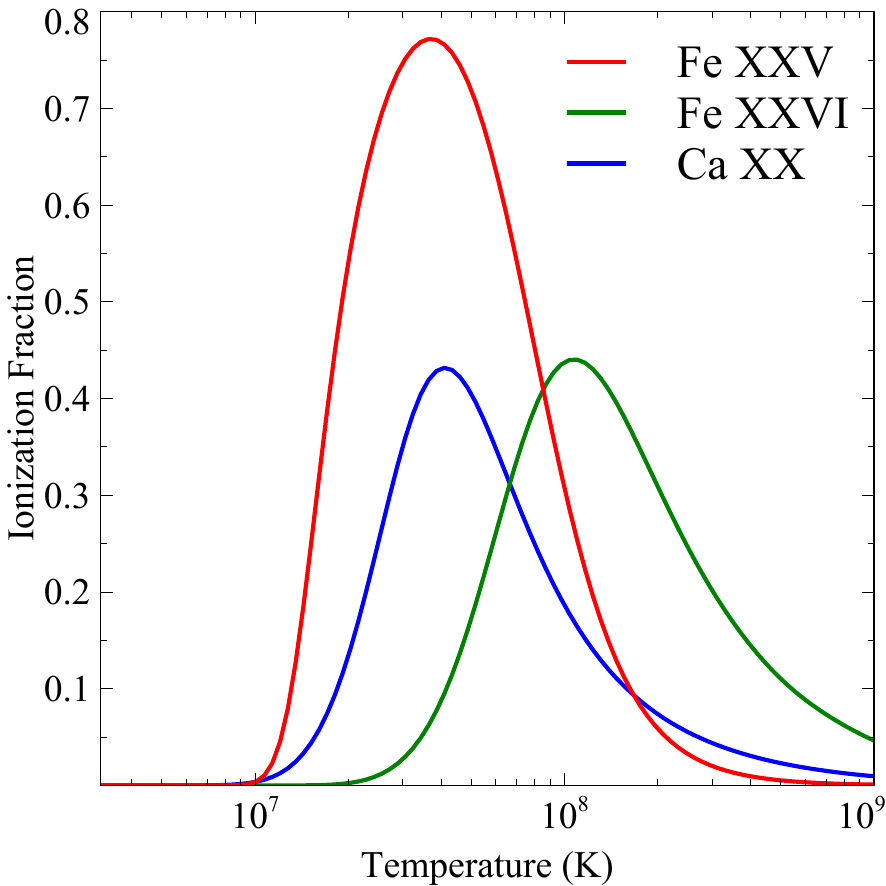}
\caption{Ionization fractions of Fe XXV, Fe XXVI, and Ca XX as a function of temperature for a collisionally ionized plasma, calculated using \texttt{AtomDB} 3.0.9. Below $T = 10^7 \, \mathrm{K}$, the ionization fractions for all three ions are negligible, indicating that collisionally ionized emission from these ions is insignificant at lower temperatures.  {Alt text: Fraction of ionization for various ions versus temperature.}
\label{f:figionfrac}}
\end{figure}

The narrow ($\sim$ 360  \text{km~s$^{-1}$}) rest-frame lines are likely originating in a photoionized outer accretion disk atmosphere, a plausible site for stationary, narrow Fe XXV and Ca XX emission \citep{2002ApJ...581.1297J}. Alternatively, these lines could be radiative recombination lines originating in the region where the accretion stream impacts the outer edge of the accretion disk, often referred to as the ``bulge." \citep{2001ApJ...557L.101C}. 
A collisionally-ionized origin for the observed narrow emission lines in the \textit{XRISM} spectrum of GX 340+0 appears less likely. The highly ionized species identified, particularly the He- and H-like stage of iron and the H-like stage of calcium exist only in extremely high temperatures (\(>10^7 \, \text{K}\)) (see Figure \ref{f:figionfrac}, derived using \texttt{AtomDB} 3.0.9; \citealt{2001ApJ...556L..91S, 2012ApJ...756..128F}). The high temperatures (\( T \sim 10^7 - 10^8 \, \mathrm{K} \)) required for collisional ionization are typically found in regions near the neutron star, such as the inner accretion disk, accretion disk corona (ADC) or boundary layer, where high turbulence, thermal motions, and gravitational effects result in significant line broadening \citep{1989ApJ...341..955K, 2001ApJ...547..355P, 2002A&A...383..524B, 2002apa..book.....F, 2010A&A...522A..96N}. The observed narrow line profiles appear inconsistent with the characteristics typically associated with collisional ionization in these environments. While jets can contribute to collisionally-ionized emission lines through shock heating or interactions with surrounding material \citep{2021MNRAS.505.5058P}, there is no evidence of jets in GX 340+0 \citep{2024A&A...691A.253L}. Therefore, we focused on modeling the spectral properties of GX340+0 using a photoionization framework.

\section*{Acknowledgments}

We acknowledge Masahiro Tsujimoto for serving as the internal reviewer of this manuscript. We acknowledge Jack Steiner for insightful discussions about the interpretation of artificially inflated iron abundances in reflection modeling.

\section*{Funding}
PC acknowledges support from NASA XRISM grant 80NSSC23K0637 and NASA Chandra grants GO3-24124X 16619325, GO3-24024X 16619349, and GO4-25094X 16619354. IP acknowledges support from
 Smithsonian Astrophysical Observatory (SAO) contract SV3-73016 to MIT for Support
of the Chandra X-Ray Center (CXC) and Science Instruments. CXC is
operated by SAO for and on behalf of NASA under contract NAS8-03060.
TY acknowledges support by NASA under award number 80GSFC24M0006. TN acknowledges support by JSPS Kakenhi 23H05441 and 23K17695.

\bibliographystyle{mnras}
\bibliography{bib}

\end{document}